\pdfoutput=1
\documentclass[a4paper,twocolumn,prb,showpacs]{revtex4}
\usepackage{amsmath}
\usepackage{amssymb}
\usepackage{graphicx}
\usepackage{verbatim}
\usepackage{color}

\newcommand{\red}[1]{\iffalse #1 \fi}
\newcommand{\KF}[1]{\iffalse #1 \fi}

\begin{document}

\title{Tunnel spectroscopy of Majorana bound states in topological superconductor-quantum
dot Josephson junctions}

\author{Guang-Yao Huang$^{1,2}$}

\author{Martin Leijnse$^{1}$}

\author{Karsten Flensberg$^{3}$}

\author{H.~Q.~Xu$^{1,4}$}

\email[Corresponding author. ]{hongqi.xu@ftf.lth.se}

\affiliation{$^{1}$Division of Solid State Physics and Nanometer Structure Consortium (nmC@LU), Lund University, Box 118,
S-221 00 Lund, Sweden}

\affiliation{$^{2}$State Key Laboratory of Optoelectronic Material and Technology
and School of Physics and Engineering, Sun Yat-Sen University, Guangzhou,
510275, China}

\affiliation{$^{3}$Center for Quantum Devices, Niels Bohr Institute, University
of Copenhagen, Universitetsparken 5, 2100 Copenhagen, Denmark}

\affiliation{$^{4}$Department of Electronics and Key Laboratory for the Physics
and Chemistry of Nanodevices, Peking University, Beijing 100871, China}

\date{\today}
\begin{abstract}
We theoretically investigate electronic transport through a junction
where a quantum dot (QD) is tunnel coupled on both sides to semiconductor
nanowires with strong spin-orbit interaction and proximity-induced
superconductivity. The results are presented as stability diagrams,
i.e., the differential conductance as a function of the bias voltage applied
across the junction and the gate voltage used to control the electrostatic
potential on the QD. A small applied magnetic field splits and modifies the
resonances due to the Zeeman splitting of the QD level. Above a critical
field strength, Majorana bound states (MBS) appear at the interfaces
between the two superconducting nanowires and the QD, resulting in
a qualitative change of the entire stability diagram, suggesting this
setup as a promising platform to identify MBS. Our calculations are
based on a nonequilibrium Green's function description and is exact
when Coulomb interactions on the QD can be neglected. In addition,
we develop a simple pictorial view of the involved transport processes,
which is equivalent to a description in terms of multiple Andreev
reflections, but provides an alternative way to understand
the role of the QD level in enhancing
transport for certain gate and bias voltages. We believe that this
description will be useful in future studies of interacting QDs coupled
to superconducting leads (with or without MBS), where it can be used
to develop a perturbation expansion in the tunnel coupling.
\end{abstract}
\maketitle

\section{Introduction}

During the last few years there has been considerable interest in
the search for Majorana bound states (MBS),\cite{np5.614} partly
because they have a unique capability for so-called topological quantum
computation.\cite{rmp80.1083} Some systems believed to host MBS include
the $\nu=\frac{5}{2}$ fractional quantum Hall state,\cite{prb61.10267}
p-wave superconductors,\cite{prl86.268} topological insulators coupled
to s-wave superconductors,\cite{prl100.096407} and two-dimensional
electron gases with strong spin-orbit interaction (SOI), coupled to
s-wave superconductors and exposed to a magnetic field.\cite{prl104.040502,prb82.214509}
A slight twist to the two-dimensional electron gas proposal is to
realize MBS in one-dimensional semiconductor nanowires with strong
SOI and large g-factors, which can be coupled to a superconductor
simply by fully or partially covering the wire with a superconducting
material.\cite{prl105.077001,prl105.177002,rpp75.076501,sst27.124003,arcmp4.113}
We will focus here on the nanowire system.

This research was originally driven by theoretical efforts, but has
very recently also attracted the attention of several experimental
groups. In order to detect the MBS which are possibly realized in
such experiments, several types of hybrid devices have been fabricated,
such as superconductor-normal metal (S-N) structures,\cite{science336.1003,np8.887}
superconductor-quantum dot-normal metal (S-QD-N) structures,\cite{nl12.6414}
and superconductor-quantum dot-superconductor (S-QD-S) structures\cite{nl12.6414,arXiv1406.4435}
(where the S electrode(s) are made from a nanowire covered with a
superconductor and (possibly) hosts the MBS). MBS can then be detected
by driving a current from one side to the other (tunnel spectroscopy),
where the presence of a MBS gives rise to a zero bias peak (ZBP) in
the conductance. Such ZBPs have indeed been experimentally observed.\cite{science336.1003,nl12.6414,np8.887,arXiv1406.4435,prb87.241401,prl110.126406}
However, ZBPs can also appear for many other reasons\cite{prl109.186802,nnano9.79,njp14.125011,prl110.217005,prb85.060507,prl109.267002}
and more evidence of MBS is needed. In devices with two superconducting
leads the $4\pi$ periodic DC Josephson effect has been theoretically
predicted to serve as a signature of MBS\cite{pu44.131} and although
this has not yet been observed experimentally, there has been reports
of unusual current-phase relations\cite{prl109.056803} and fractional
AC Josephson effect\cite{np8.795} which might also indicate the presence
of MBS. 
Calculations have shown that in a topological weak link between two trivial superconductors the 
existence of MBS changes the subgap features related to multiple Andreev reflection (MAR)\cite{Badiane11}.
It has also been shown theoretically that MAR in a weak link between two superconductors in the
topological phase (i.e., hosting MBS) could cause novel subgap structures
different from the trivial case,\cite{njp15.075019} which can also
be regarded as signatures of MBS.

In this paper, we investigate a S-QD-S setup. Nanowires as links between
two superconducting leads were experimentally realized almost a decade
ago~\cite{science309.272,nature442.667} (although these studies
did not aim to realize MBS) and it was shown that the supercurrent
can be controlled by a gate voltage. We investigate instead a voltage-biased
junction and, assuming that the QD level can be controlled by a gate
voltage, we calculate the full stability diagram, i.e., the nonlinear
differential conductance as a function of gate and bias voltage. To
this end, we calculate the time-averaged AC Josephson current and
differential conductance ($dI/dV_{b}$) using the nonequilibrium Green's
function (NEGF) method. We show that the stability diagram looks completely
different when the superconducting electrodes host MBS compared to
the case without MBS. This allows detection of MBS through the qualitative
appearance of the entire stability diagram, rather than just from
a zero-bias peak which is more likely to arise for other reasons.
To complement the calculations, we develop a simple pictorial view
of multiple Andreev reflection (MAR) in S-QD-S junctions and analyze
and explain how this changes when MBS are present.

The paper is structured in the following way. In Sec.~II, we introduce
our setup and model. The leads can be driven into a topological phase
with MBS appearing at the edges next to the QD, giving rise to novel
Majorana-related signals in the conductance. The details of the calculation
can be found in the Appendix. The main results are presented in Sec.~III,
where we show the calculated stability diagrams with and without
MBS and analyze the positions of the peaks in both cases. The main
focus is on the limit of weak S-QD tunnel coupling, but we also present
results in the strong-coupling limit. Section IV summarizes and concludes.

\section{Model and method}

\begin{figure}
\centering\includegraphics[scale=0.5]{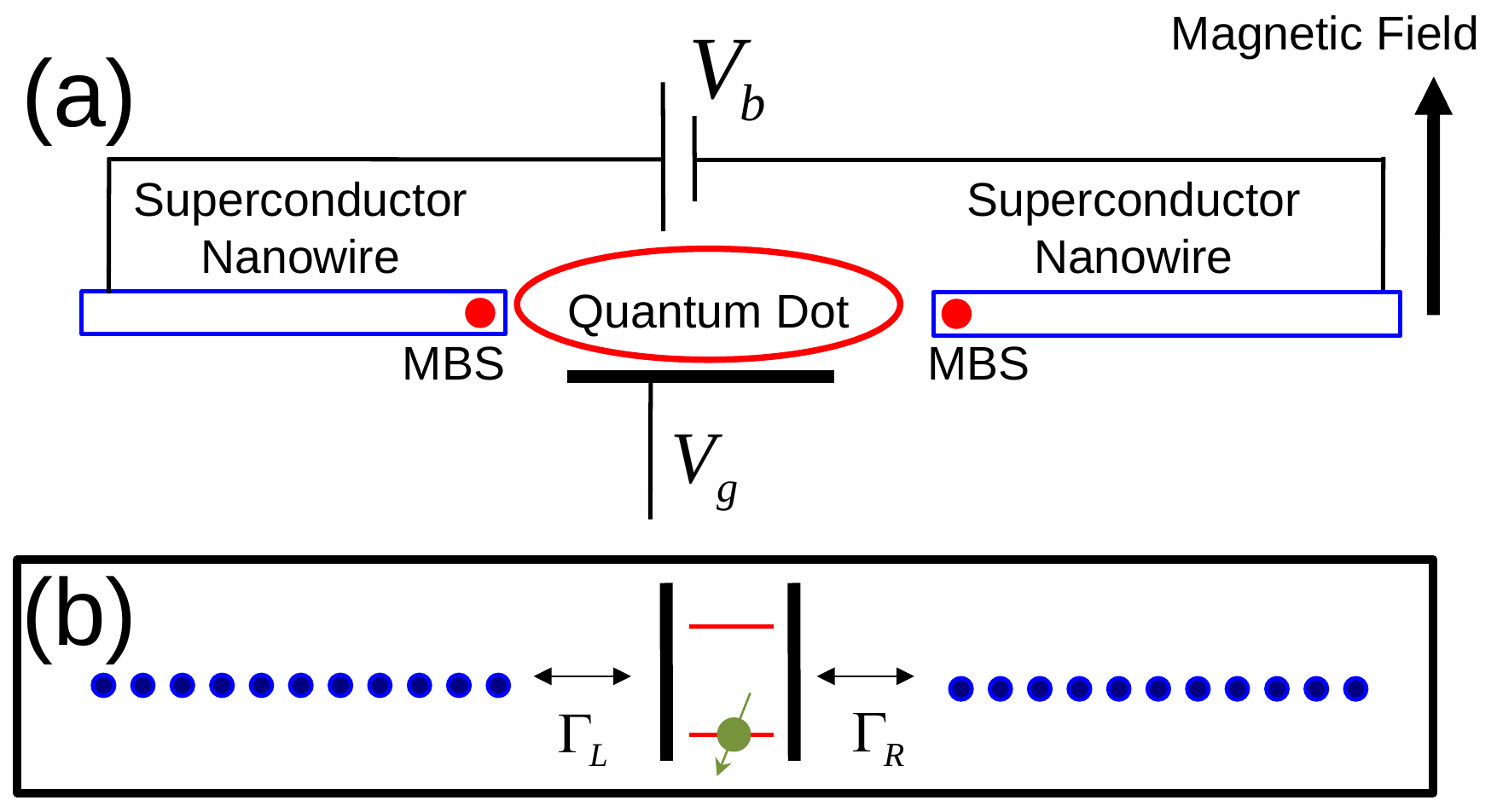}
\caption{\label{setup} (a) Sketch of the transport setup with a voltage-biased
nanowire S-QD-S configuration. The bias voltage $V_{b}$ is applied
between the superconducting contacts and the electrostatic potential
on the QD is controlled by the gate voltage $V_{g}$. (b) The model
abstracted from the setup in (a). The leads are semi-infinite tight-binding
chains with the end sites coupling to the QD with tunnel couplings
$\Gamma_{L}$ and $\Gamma_{R}$.}
\end{figure}

In our S-QD-S transport setup (see Fig.~\ref{setup}), the total Hamiltonian
$H_{total}$ consists of three parts: the leads $H_{\alpha=L,R}$,
the QD $H_{QD}$ and the coupling between them $H_{c}$.

The leads and the QD are realized in a one-dimensional nanowire along
the $x$-axis with strong SOI which is exposed to a magnetic field.
The leads have been made superconducting by proximity-coupling to
s-wave Bardeen-Cooper-Schrieffer (BCS) superconductors (not shown).
Introducing the electron creation/annihilation operator $c_{n\sigma}^{\dagger}/c_{n\sigma}$
for site $n$ and spin $\sigma$, after discretizing the continuous
Hamiltonian,\cite{prl105.077001,prl105.177002} the tight-binding
Hamiltonian is $H_{\alpha}=\sum_{nm}H_{n,m}^{\alpha}$, where
\begin{eqnarray}
H_{n,m}^{\alpha} & = & \sum_{\sigma}c_{n\sigma}^{\dagger}\delta_{n,m}[(2t-\mu)+V_{z}\sigma_{z}]_{\sigma\sigma}c_{m\sigma} \nonumber \\
 & + & \sum_{\sigma\sigma'}c_{n\sigma}^{\dagger}[(-t+i\alpha_{0}\sigma_{y})\delta_{n,m-1}+h.c.]_{\sigma\sigma'}c_{m\sigma'}\nonumber \\
 & + & \Delta\delta_{n,m}(e^{i\phi}c_{n\downarrow}^{\dagger}c_{m\uparrow}^{\dagger}+h.c.) .\label{eqHLead}
\end{eqnarray}
Here $m,n$ are the site indices, $t=\frac{\hbar^{2}}{2m^{*}a^{2}}$
is the parameter related to band width, with $\hbar$ the reduced
Planck constant, $m^{*}$ the effective mass, and $a$ the lattice
spacing. $\alpha_{0}$ is the Rashba SOI strength, $\mu$ is the chemical
potential, $V_{z}$ is the Zeeman energy, and $\Delta$ and $\phi$
are the absolute value and phase of the superconductor pair potential,
respectively. In Eq. (\ref{eqHLead}) and below we have suppressed
the lead index $\alpha$ on all quantities even though they may be
different in the two leads (in the actual calculations presented,
only $\mu_{\alpha}$ are different for $\alpha=L$ and $\alpha=R$).
Rewritten in Nambu basis
\begin{equation}
C_{n}=\left[\begin{array}{cccc}
c_{n\uparrow} & c_{n\downarrow} & c_{n\downarrow}^{\dagger} & -c_{n\uparrow}^{\dagger}\end{array}\right]^{T},
\end{equation}
each lead is described by a tight-binding Bogoliubov-de Gennes (BdG)
Hamiltonian
\begin{eqnarray}
H_{\alpha} & = & \frac{1}{2} \sum_{n,m}C_{n}^{\dagger}(\mathcal{H}_{BdG}^{\alpha})_{n,m}C_{m}+\mathrm{const.},\\
(\mathcal{H}_{BdG}^{\alpha})_{n,m} & = & h_{0}\delta_{n,m}+h_{1}\delta_{n,m-1}+h_{-1}\delta_{n,m+1},
\end{eqnarray}
where
\begin{eqnarray}
h_{0} & = & (2t-\mu)\tau_{z}+V_{z}\sigma_{z}+\Delta(\tau_{x}\cos\phi-\tau_{y}\sin\phi),\\
h_{1} & = & (-t+i\alpha_{0}\sigma_{y})\tau_{z},\\
h_{-1} & = & (-t-i\alpha_{0}\sigma_{y})\tau_{z}.
\end{eqnarray}
The Pauli matrices $\sigma_{i=x,y,z}$, $\tau_{i=x,y,z}$ operate
on spin and particle-hole spaces, respectively. Note that $\Delta$
is the induced superconducting pairing potential in the nanowire,
which is experimentally found in
InSb and InAs wires to be in the range $0.13-0.45$~meV,\cite{np8.887,nl12.6414,arXiv1406.4435,np8.795,nl12.228,nl12.5622,nl13.3614,prb89.214508}
while the SOI strength $\alpha_{0}$ typically is $0.07-0.3$~meV.\cite{np8.887,nl9.3151}

The applied bias voltage $V_{b}$ acting on the superconductor lead
enters $\mathcal{H}_{BdG}^{\alpha}$ through a change of chemical
potential, which can be transferred to a time-dependent phase of the
Nambu basis.\cite{prb61.4754,prl107.256802} The phase of the superconductor
pair potential $\phi$ can also be removed from $\mathcal{H}_{BdG}^{\alpha}$
to the Nambu basis through a similar gauge transformation. These transformations
result in a time-dependent coupling of the QD and lead in Eq.~(\ref{eqCouplingHamiltonian}).
The relevant lead Green's function is that of the site closest to
the QD, which can be found numerically by extending the lead tight-binding
chain to infinity.\cite{prb66.165305,prb72.195346,Wimmer} This semi-infinite lead Green's function
captures both bulk states and possible edge states (such as MBS). Note that we here assume the potential resulting from the bias voltage to drop only at the tunnel barriers defining the QD. The possibility that, e.g., surface roughness causes the bias voltage to drop over the whole nanowire and form a Majorana island will be considered elsewhere.

The single-level QD is described by the Hamiltonian
\begin{equation}
H_{QD}=\sum_{\sigma}d_{\sigma}^{\dagger}\left(E_{0}-eV_{g}+V_{z}\sigma_{z}-\frac{eV_{b}}{2}\right)_{\sigma\sigma}d_{\sigma},\label{QDHam}
\end{equation}
where $d_{\sigma}^{\dagger}/d_{\sigma}$ is the creation/annihilation
operator of the QD, $E_{0}$ is the energy of the QD level without
applied voltages, and $V_{g}$ is the gate voltage (for simplicity
we set the gate coupling to one). The Zeeman energy $V_{z}$ is assumed
to be the same as in the lead. The term involving $V_{b}$ appears
because we assume the bias to be applied only to the right lead, $V_{R}=-V_{b}$,
$V_{L}=0$, and assume the capacitances associated with the right
and left lead to be equal and much larger than the gate capacitance
(the physics is equivalent to using symmetric bias, $V_{R}=-V_{b}/2$,
$V_{L}=+V_{b}/2$, with a QD level independent of $V_{b}$, but for
technical reasons it is easier to consider only one lead to be biased).
Rewritten in Nambu basis
\begin{equation}
D=\left[\begin{array}{cccc}
d_{\uparrow} & d_{\downarrow} & d_{\downarrow}^{\dagger} & -d_{\uparrow}^{\dagger}\end{array}\right]^{T}
\end{equation}
the Hamiltonian becomes
\begin{equation}
H_{QD}=\frac{1}{2}D^{\dagger}\mathcal{H}_{BdG}^{QD}D+\mathrm{const.},
\end{equation}
with
\begin{equation}
\mathcal{H}_{BdG}^{QD}=(E_{0}-eV_{g}-\frac{eV_{b}}{2})\tau_{z}+V_{z}\sigma_{z}.
\end{equation}
The retarded Green's function of the isolated QD is
\begin{equation}
g_{QD}^{r}(\omega)=(\omega-\mathcal{H}_{BdG}^{QD}+i\delta)^{-1}
\end{equation}
and the advanced Green's function is $g_{QD}^{a}(\omega)=g_{QD}^{r\dagger}(\omega)$,
where $\delta=0^{+}$ is an infinitesimal positive number (in the
numerical calculations we will use a small finite $\delta$).

Tunneling between the QD and leads is described by the coupling Hamiltonian,
which in terms of the Nambu basis defined above is given by
\begin{equation}
H_{c}=\frac{1}{2}\sum_{\alpha k}[C_{\alpha k}^{\dagger}T_{\alpha}(\tau)D+D^{\dagger}T_{\alpha}^{\dagger}(\tau)C_{\alpha k}],\label{eqCouplingHamiltonian}
\end{equation}
where $T_{\alpha}(\tau)=t_{c}\tau_{z}e^{i\tau_{z}eV_{\alpha}\tau}$.
The coupling $t_{c}$ is assumed to be real, independent of lead momentum,
and to only couple components with the same spin. With the bias being
applied only to the right lead, the couplings are $T_{L}(\tau)=T_{L}=t_{c}\tau_{z}$,
$T_{R}(\tau)=t_{c}\tau_{z}e^{-i\tau_{z}eV_{b}\tau}$.

The time-averaged current $I_{DC}$ is (see the Appendix for a derivation)
\begin{eqnarray}
I_{DC} & = & \frac{e}{h}\mathrm{Re}\mathrm{Tr}\int_{-eV_{b}}^{eV_{b}}d\omega\nonumber \\
 & \times & [\boldsymbol{G}_{QD}^{r}(\omega)\boldsymbol{\Sigma}_{L}^{<}(\omega)+\boldsymbol{G}_{QD}^{<}(\omega)\boldsymbol{\Sigma}_{L}^{a}(\omega)]\tau_{z},\label{eqIDC}
\end{eqnarray}
where $\boldsymbol{G}$ and $\boldsymbol{\Sigma}$ are matrices in
Fourier space, and the trace is also taken in Fourier space.\cite{prb78.024518}
The full retarded Green's function of the QD is derived from the Dyson
equation $\boldsymbol{G}_{QD}^{r}=\boldsymbol{g}_{QD}^{r}+\boldsymbol{G}_{QD}^{r}\boldsymbol{\Sigma}^{r}\boldsymbol{g}_{QD}^{r}$,
\begin{equation}
\boldsymbol{G}_{QD}^{r}=[(\boldsymbol{g}_{QD}^{r})^{-1}-\boldsymbol{\Sigma}^{r}]^{-1},
\end{equation}
and the lesser Green's function is obtained from the Keldysh equation
\begin{equation}
\boldsymbol{G}_{QD}^{<}=\boldsymbol{G}_{QD}^{r}\boldsymbol{\Sigma}^{<}\boldsymbol{G}_{QD}^{a},
\end{equation}
where retarded/lesser self energies are $\boldsymbol{\Sigma}^{r/<}=\boldsymbol{\Sigma}_{L}^{r/<}+\boldsymbol{\Sigma}_{R}^{r/<}$.
The Fourier components of the above Green's functions and self energies
are
\begin{eqnarray}
 &  & g_{QD,mn}^{r}(\omega)=\delta_{mn}g_{QD}^{r}(\omega_{m}),\\
 &  & \Sigma_{L,mn}^{r/<}(\omega)=\delta_{mn}\Gamma\tau_{z}g_{L}^{r/<}(\omega_{m})\tau_{z},\label{eqSelfEnergyL}\\
 &  & \Sigma_{R,mn}^{r/<}(\omega)=\Gamma\\
 & \times & \left[\begin{array}{cc}
\delta_{mn}g_{R,11}^{r/<}(\omega_{m+\frac{1}{2}}) & -\delta_{m,n-1}g_{R,12}^{r/<}(\omega_{m+\frac{1}{2}})\\
-\delta_{m,n+1}g_{R,21}^{r/<}(\omega_{m-\frac{1}{2}}) & \delta_{mn}g_{R,22}^{r/<}(\omega_{m-\frac{1}{2}})
\end{array}\right],\nonumber
\end{eqnarray}
where the coupling strength between the QD and both leads is the same,
$\Gamma=t_{c}^{2}$, and subscript $ij$ in $g_{R,ij}^{r/<}(\omega)$
means the $ij-$th 2$\times$2 block of the corresponding 4$\times$4
matrix. The lesser Green's function of the lead is related to the
retarded Green's function as
\begin{eqnarray}
g_{L/R}^{<}(\omega) & = & -f(\omega)[g_{L/R}^{r}(\omega)-g_{L/R}^{a}(\omega)]\\
 & = & -2if(\omega)\mathrm{Im}g_{L/R}^{r}(\omega),
\end{eqnarray}
where $f(\omega)=1/(e^{\frac{\omega}{k_{B}T}}+1)$ is the Fermi-Dirac
distribution function with $k_{B}$ the Boltzmann constant and $T$
the temperature. The second equality is a result of the lead Green's
function being symmetric when $\phi=0$, which can always be satisfied
by a gauge transformation.

The current is calculated from Eq.~(\ref{eqIDC}) by numerical integration
and the size of the matrices in Fourier space is increased until the
current has converged.

\section{Results and analysis}

\subsection{Density of states at the end of the leads}

\begin{figure}
\centering\includegraphics[scale=0.6]{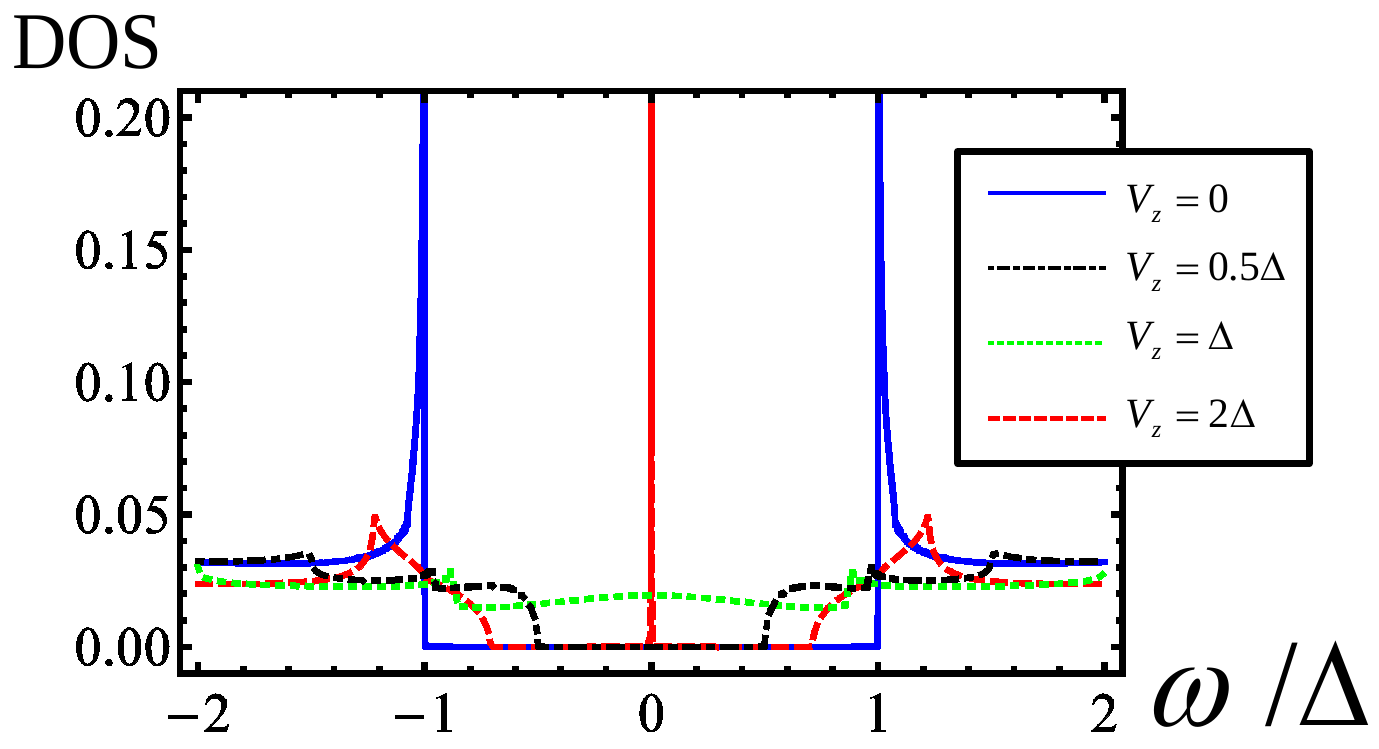}
\caption{\label{figDOS} DOS at the end site of an isolated lead for $t=10\Delta$, $\mu=0$, $\alpha_{0}=2\Delta$, and $\delta=10^{-5} \Delta$
in the retarded Green's function. The different curves are the results for
Zeeman energy $V_{z}=0$ (blue line), $V_{z}=0.5\Delta$ (black dot-dashed
line), $V_{z}=\Delta$ (green dotted line), and $V_{z}=2\Delta$ (red dashed line).}
\end{figure}

We first examine the density of states (DOS) at the end site of an
isolated lead, which is calculated from the Green's function
\begin{equation}
\rho(\omega)=-\frac{1}{\pi}\mathrm{Tr}\left\{ \mathrm{Im}g_{L}^{r}(\omega)\right\} .
\end{equation}
For $V_{z}=0$ the DOS exhibits the well known BCS singularities and
superconducting gap, see blue line in Fig. \ref{figDOS}. As the Zeeman
energy $V_{z}$ increases, the energy gap $E_{g}$ decreases from
$E_{g}(V_{z}=0,\mu=0)\equiv\Delta$ and the singularities at the edge of
energy gap become smoother, see black dot-dashed line in Fig.~\ref{figDOS}
where $V_{z}=0.5\Delta$. The gap closes  when $V_{z}=\sqrt{\Delta^{2}+\mu^{2}}$, shown
in the green dotted line in Fig.~\ref{figDOS}.
For larger $V_{z}$ the gap opens again, but the superconductor is
now in a topological phase and a MBS emerges as a sharp peak in the DOS,
which persists at zero energy regardless of how the Zeeman energy
varies, see the red dashed line in Fig.~\ref{figDOS}.

\subsection{Tunnel spectroscopy in the trivial phase without MBS}

In order to have a clear comparison, we first investigate tunnel spectroscopy
in a S-QD-S junction without MBS. We focus on the tunneling limit,
where the coupling is weak enough that only tunneling processes of
low order in $\Gamma$ are visible in the differential conductance
$dI/dV_{b}$ (which is fulfilled when $\Gamma / \Delta < 1$).

It is well-known that a junction between two superconductors with
not too weak coupling can exhibit subgap structures inside the gap
at $eV_{b}=\frac{2E_{g}}{m}$. This was explained in terms of MAR
by Octavio et al. (OBTK model),\cite{prb27.6739,prb38.8707} using
an incoherent Boltzman equation approach. In coherent superconducting
junctions multi-particle tunneling has been considered in terms of perturbation
theory in the tunneling coupling \cite{prl10.17,prb9.3757} and later
theories \cite{jltp68.1,prl74.2110,prb54.7366,prl75.1831} further
increased the understanding of MAR, with application to for example
quantum point contacts\cite{nl12.228} and resonant structures, such
as S-N-S junctions\cite{prb64.144504,prb60.1382}, molecules\cite{prb73.214501} or single level QDs.\cite{prb55.6137,prb65.075315,prb80.184510,prl91.057005,prb65.180514,prb86.235427}
In a S-QD-S structure, the QD levels shift the peaks of MAR considerably.
It has been argued that the subgap structures in S-QD-S devices can be understood in terms of enhancements
when the QD level lies on the trajectory of a MAR process. Here, we
present an alternative picture which only relies on energy
conservation and which more clearly reveals why some MAR resonances
are enhanced and some are suppressed in the presence of a QD. Below
this is first used to explain our calculations for the S-QD-S junction
in the situation without MBS.

\begin{figure*}[htb!]
\centering\includegraphics[scale=0.9]{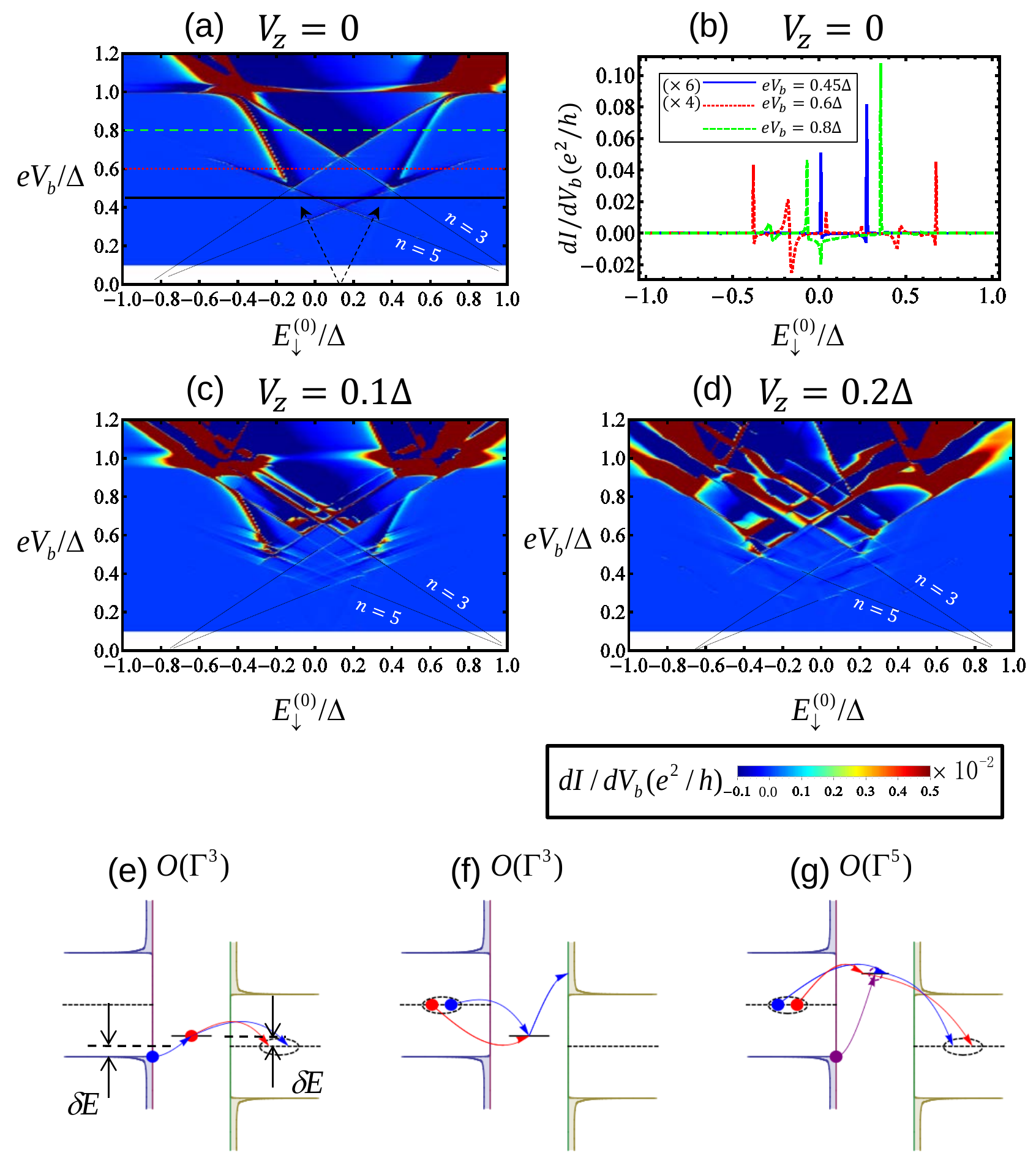}
\caption{\label{MARTrivial} $dI/dV_{b}$ as a function of $eV_{b}$ and $E_{\downarrow}^{(0)}$
for (a) Zeeman energy $V_{z}=0$, (c) $V_{z}=0.1\Delta$,
and (d) $V_{z}=0.2\Delta$. The color scale is limited to the range
$[-0.001,0.005]$$e^{2}/h$. In (a), (c) and (d), the dotted black lines, added as guides
to the eye, are given by $eV_{b}=\frac{2}{3}(E_{g}\pm(E_{\downarrow}^{(0)}-\delta_{\Gamma}))$ with a mark $n=3$
and $eV_{b}=\frac{2}{5}(E_{g}\pm(E_{\downarrow}^{(0)}-\delta_{\Gamma}))$ with a mark $n=5$ next to the peaks,
and the black arrows point at the resonances given by Eq.~(\ref{CPresonance}).
Note that $E_{g}$ is the energy gap which is also changed when $V_{z}$
varies. We do not show results for $eV_{b}<0.1\Delta$ because the
number of harmonics which have to be taken into account when evaluating
the Green's functions grow with decreasing $V_{b}$. We have chosen
$\Gamma=0.9\Delta$ and $k_{B}T=0.01\Delta$. $\delta_{\Gamma}=0.13\Delta$
was found to give the best fit to the peaks. All other parameters
are the same as in Fig. \ref{figDOS}.
(b) Shows $dI/dV_{b}$ as a function of  $E_{\downarrow}^{(0)}$ for $V_{z}=0$, i.e., horizontal cuts in (a),
for $eV_b=0.45\Delta$ (blue line, multiplied by 6), $0.6\Delta$ (red dotted line, multiplied by 4) and $0.8\Delta$ (green dashed line).
(e) Schematic diagram of an
$O(\Gamma^{3})$ tunnel process at $eV_{b}=0.8\Delta$ and $E_{\downarrow}^{(0)}=-0.2\Delta$
which empties an initially full QD. (f) Same as (e), but showing a
process filling an initially empty QD with one electron. (g) Same
as (e), but showing an $O(\Gamma^{5})$ tunnel process at $E_{\downarrow}^{(0)}=\Delta$.}
\end{figure*}

In Fig.~\ref{MARTrivial} (a), (c), and (d), we show $dI/dV_{b}$ as a function of $eV_{b}$
and $E_{\downarrow}^{(0)}$, the latter being the energy of a spin-down
electron on the QD, i.e., $E_{\sigma=\uparrow,\downarrow}^{(0)}=E_{0}-eV_{g}\pm V_{z}$,
without the effect of bias.
Figure~\ref{MARTrivial} (b) shows $dI/dV_{b}$ as a function of  $E_{\downarrow}^{(0)}$ for $V_{z}=0$ and
different values of $eV_b$, i.e., horizontal cuts in Fig.~\ref{MARTrivial} (a).
We first note that we recover previously
discussed features of transport in S-QD-S junctions, such as the existence
of negative differential conductance \cite{prb55.6137} and the absence
of all the even order MAR peaks along $E_{\downarrow}^{(0)}=0$. \cite{prb80.184510,prl91.057005}
There are two different types of peaks. The first type, which is marked with black arrows in Fig.~\ref{MARTrivial}(a),
appears along the lines
\begin{equation}
eV_{b}=\pm2\bar{E},\label{CPresonance}
\end{equation}
where $\bar{E}=(E_{\uparrow}^{(0)}+E_{\downarrow}^{(0)})/2$ is the
average energy of the spin split QD level. When $V_{z}=0$, this is
the same as the position of the edges of Coulomb diamonds in transport
through QDs coupled to normal leads, but in the case of superconducting
leads it is related to the possibility for a Cooper pair to tunnel
onto or off the QD. Moreover, in contrast to the standard Coulomb
diamond edges these peaks are not split in a magnetic field,
because of the singlet nature of the Cooper pairs.

The second type of peak appears along the lines where $eV_{b}$ and
$E_{\sigma}^{(0)}$ approximately satisfy
\begin{equation}
eV_{b}=\frac{2}{n}(E_{g}\pm E_{\sigma}^{(0)}),\label{resonance}
\end{equation}
where $n=3,5,7,\ldots$ (only the lines corresponding to $n=3,5$
are visible in Fig.~\ref{MARTrivial} and are marked with dotted lines in Fig.~\ref{MARTrivial}(a), (c) and (d)). Notice, however, that these
lines reduce to $eV_{b}=\frac{E_{g}}{m}$, with $n=2m\pm 1$ at the resonance
conditions $E_{\sigma}^{(0)}=\frac{E_{g}}{2m}$, in agreement with the picture
that the MAR path goes through that QD level.

To understand the appearance of these lines, consider the lowest order
(in $\Gamma$) tunnel process which is allowed, meaning that energy
is conserved in the entire process. If the QD level is within the
gap, tunneling into or out of the QD with a single electron (tunnel
rate $\propto\Gamma$) can never conserve energy. Figure~\ref{MARTrivial}(e)
shows the lowest order tunnel process ($\propto\Gamma^{3}$) which
empties an initially singly occupied QD level inside the gap while
conserving energy: the electron on the QD tunnels into the right lead
(red arrow, $\propto\Gamma$), where it forms a Cooper pair together
with an electron which cotunnels from the left to the right lead (blue
arrows, $\propto\Gamma^{2}$). If the QD level lies at $\delta E$
above the chemical potential of the right lead [see Fig.~\ref{MARTrivial}(e)],
the electron orginally residing on the QD gains the energy $\delta E$
when forming a Cooper pair, which must be compensated by taking the
second electron from a quasiparticle state in the left lead at $\delta E$
below the chemical potential of the right lead. Existence of such
quasiparticle state requires that $eV_{b}>\frac{2}{3}(E_{g}-E_{\sigma}^{(0)})$.
For a stationary current to flow, it must also be possible to fill
the QD with an electron again and the corresponding lowest order process
($\propto\Gamma^{3}$) is shown in Fig.~\ref{MARTrivial}(f): a Cooper
pair breaks up in the left lead, with one electron tunneling onto
the QD (red arrow, $\propto\Gamma$) and the other cotunneling through
the QD into the quasiparticle states above the gap in the right lead
(blue arrows, $\propto\Gamma^{2}$). Energy conservation gives that
this process is possible when $eV_{b}>\frac{2}{3}(E_{g}+E_{\sigma}^{(0)})$.
These conditions give peaks in $dI/dV_{b}$ according to Eq.~(\ref{resonance})
with $n=3$. For $E_{\sigma}^{(0)}=0$ the two conditions are the
same and identical to the condition for third order MAR in a junction
without a QD, $eV_{b}>\frac{2E_{g}}{3}$. Thus, the presence of the
QD level enhances the current by allowing a third order MAR process
(which is $\propto\Gamma^{6}$) to be split into two consecutive processes
$\propto\Gamma^{3}$, involving real (rather than virtual) occupation
of the QD. Similar arguments show that Eq.~(\ref{resonance}) in
general corresponds to the onset of tunnel processes $\propto\Gamma^{n}$.
An example with $n=5$ is shown in Fig.~\ref{MARTrivial}(g),
where a Cooper pair cotunnels through the QD (red and blue arrows, $\propto\Gamma^4$) from the left to the right lead while one electron tunnels into the QD from quasiparticle states below the gap in the left lead (purple arrow, $\propto\Gamma$).
In general, the QD level allows a MAR process to be split into two separate
tunnel processes of lower order in $\Gamma$. We have thus developed
a perturbative (in $\Gamma$) way of understanding the observation
that tunneling is enhanced when the QD level lies on the path of a
MAR process.\cite{prb55.6137,prb80.184510,prl91.057005}

Upon closer inspection we see that the peaks do not exactly fit Eq.~(\ref{resonance}),
but are shifted by a more or less voltage-independent energy $\delta_{\Gamma}$.
The origin of this shift is tunneling renormalization of the QD level
position,\cite{prb77.161406} i.e., $E_{\sigma}^{(0)}\to E_{\sigma}^{(0)}-\delta_{\Gamma}$.
The renormalization effect is much stronger here than with standard
BSC superconducting leads because the chemical potential is close
to the bottom of the band, creating a strong energy asymmetry in the
number of available lead states. In Fig.~\ref{MARTrivial}(a) we
have included the shift in the dashed lines indicating the resonance
positions. \KF{Should we quantified this further?}

With a finite Zeeman energy $V_{z}$, see Fig.~\ref{MARTrivial}(c)--\ref{MARTrivial}(d),
the breaking of the spin degeneracy of the QD level results in a splitting
of these resonances because $E_{\uparrow}^{(0)}-E_{\downarrow}^{(0)}=2V_{z}$.
The increased smoothness of the DOS and decreased $E_g$ as $V_{z}$ increases
(see the DOS for $V_{z}=0.5$ in Fig.~\ref{figDOS}) further modifies the peaks.

\subsection{Tunnel spectroscopy in the topological phase with MBS}

\begin{figure*}[htb!]
\centering\includegraphics[scale=0.9]{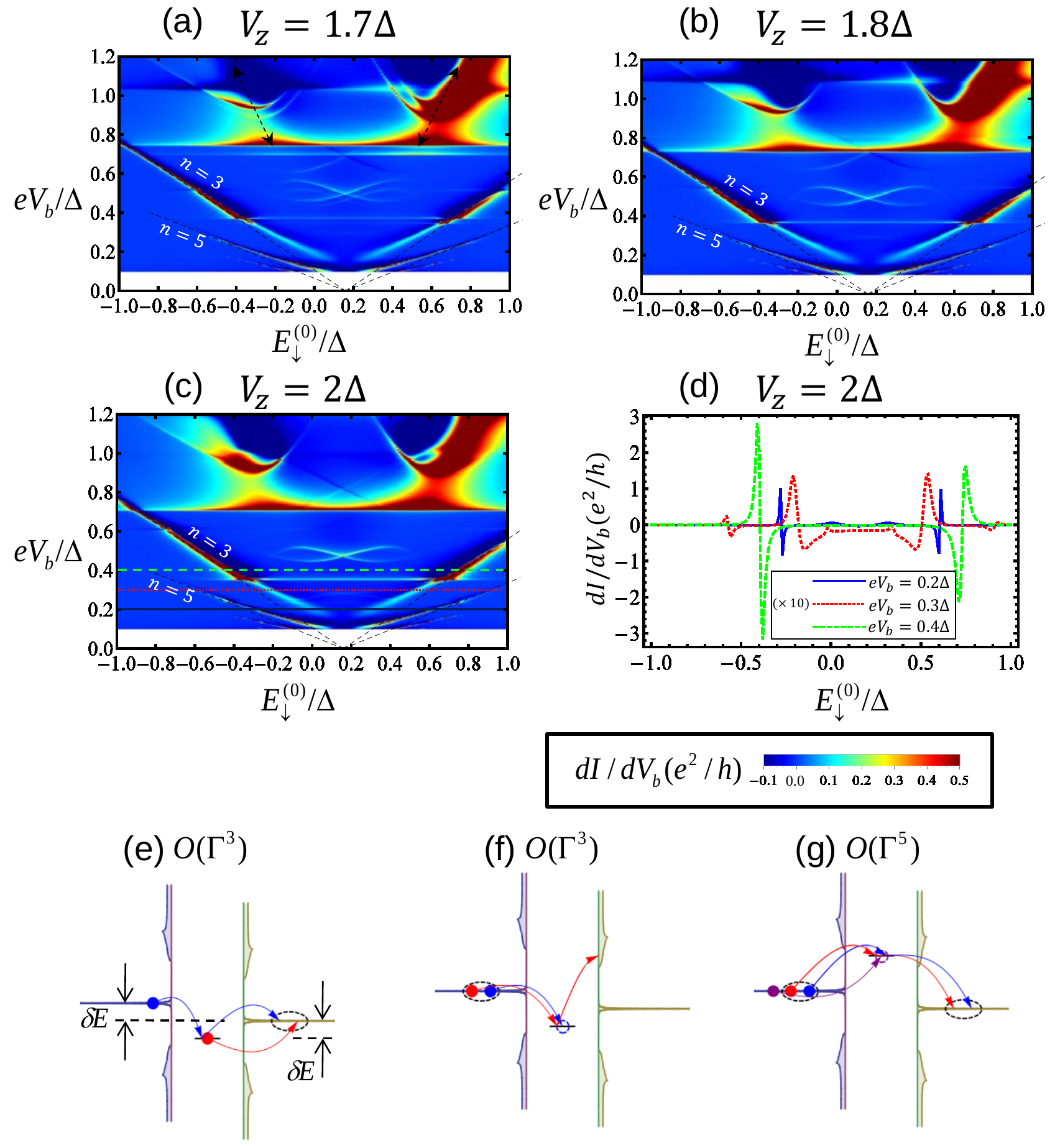}
\caption{\label{MARMFQDMF}$dI/dV_{b}$ as a function of $eV_{b}$ and $E_{\downarrow}^{(0)}$
with a large Zeeman energy, $V_{z}=1.7\Delta$ in (a), $V_{z}=1.8\Delta$
in (b), and $V_{z}=2.0\Delta$ in (c). The color scale is limited to the range
$[-0.1,0.5]e^{2}/h$. The dashed black lines are guides to the eye
given by $eV_{b}=\pm\frac{2}{3}(E_{\downarrow}^{(0)}-\delta_{\Gamma})$ marked with $n=3$
and $eV_{b}=\pm\frac{2}{5}(E_{\downarrow}^{(0)}-\delta_{\Gamma})$ marked with $n=5$,
where $\delta_{\Gamma}=0.16\Delta$.
(d) Shows $dI/dV_{b}$ as a function of  $E_{\downarrow}^{(0)}$ for $V_{z}=2\Delta$, i.e., horizontal cuts in (c),
for $eV_b=0.2\Delta$ (blue line), $0.3\Delta$ (red dotted line, multiplied by 10)
and $0.4\Delta$ (green dashed line).
We have used $\delta = 10^{-3}$, the other parameters are the same as in Fig.~\ref{MARTrivial}.
(e)--(g): Schematic diagrams of tunnel processes in the presence of
MBS corresponding to (c) at $eV_{b}=0.3\Delta$. In (e) and (f) $E_{\downarrow}^{(0)}=-0.45\Delta$,
while $E_{\downarrow}^{(0)}=0.75\Delta$ in (g). We have for illustrative purposes drawn electrons inside the
MBS peaks, although in reality the charge will disappear into the superconductor and not be localized at the edge.}
\end{figure*}

With our detailed understanding of the transport features of the S-QD-S
junction in the trivial phase, we are now ready to consider the case
with MBS by tuning the magnetic field to drive the leads into the
topological superconducting phase, which is the central result in
this paper. The huge difference in the stability diagram when comparing
the trivial phase and the topological phase with MBS arises from the
possibility, introduced by the MBS, to tunnel into the leads with
single electrons exactly in the middle of the gap. In addition, the
smoothening of the singularities at the edge of the superconducting
gap and the spin polarization due to the large Zeeman splitting of
the QD levels result in a suppression of the conventional MAR resonances,
making the MBS related transport signatures stand out in the stability
diagram.

Typical stability diagrams in the topological phase are shown in Fig.
\ref{MARMFQDMF}(a)--\ref{MARMFQDMF}(c), and  Fig.~\ref{MARMFQDMF}(d) shows
$dI/dV_b$ as a function of $E_\downarrow^{(0)}$ at constant $eV_b$.
There are two new types of peaks not present
without MBS. The first type appears at
\begin{equation}
eV_{b}=2E_{\sigma}^{(0)},\; eV_{b}>E_{g},\label{Mresonance1}
\end{equation}
and are marked with black arrows in Fig. \ref{MARMFQDMF}(a).
(As in the trivial case, tunneling renormalization shifts the QD level,
$E_{\sigma}^{(0)}\to E_{\sigma}^{(0)}-\delta_{\Gamma}$, which
we for simplicity ignore in the resonance conditions.) Note that these
lines are different from those in the trivial case described by Eq.~(\ref{CPresonance})
since they do not correspond to coherent tunneling of Cooper pairs,
but rather to processes where single electrons tunnel directly from
the Fermi level in one lead (made possible by the MBS) onto the QD and then out to the quasi-particle
states outside the gap in the other lead, or vice versa. When Eq.~(\ref{Mresonance1}) is fulfilled, these processes
involve resonant tunneling into/out of the QD level and the rate
is $\propto\Gamma$, similar to standard sequential tunneling, giving
rise to the very large conductance. Note also that $E_{\sigma}^{(0)}$
rather than $\bar{E}$ appears in Eq.~(\ref{Mresonance1}) and the
peaks are completely missing for $eV_{b}<E_{g}$.

The second type of new peaks appear at
\begin{equation}
eV_{b}=\pm\frac{2}{n}E_{\sigma}^{(0)},\label{Mresonance2}
\end{equation}
where $n=3,5,7,\ldots$ corresponds to tunnel processes $\propto\Gamma^{n}$
(only the peaks with $n=3,5$ are visible in Fig.~\ref{MARMFQDMF}(a)--\ref{MARMFQDMF}(c) and are marked with dashed
lines).
These peaks can be understood in the same way as without MBS, except
that single electrons can now tunnel into or out of the MBS inside
the gap.

Let us focus on $E_{\downarrow}^{(0)}<0$ and try to understand the
$n=3$ line with negative slope. An initially occupied QD can be emptied
as shown in Fig. \ref{MARMFQDMF}(e): one electron tunnels from the
QD into the right lead (red arrow, $\propto\Gamma$) where it forms a Cooper
pair together with an electron cotunneling from the Fermi energy in the left
lead (blue arrows, $\propto\Gamma^{2}$). The energy $\delta E$ gained by the
first electron in forming a Cooper pair has to be compensated by the
energy lost by the second electron. This is possible when $eV_{b}=-\frac{2}{3}E_{\downarrow}^{(0)}$.
Note that since no quasiparticle states are involved, this process
is allowed only at this $V_{b}$, not at higher voltages. It is clear
from Fig.~\ref{MARMFQDMF}(a)--\ref{MARMFQDMF}(c) that this peak becomes very weak
for small $V_{b}$, which is related to processes filling the QD again.
A process $\propto\Gamma^{3}$ filling the QD is shown in Fig. \ref{MARMFQDMF}(f):
A Cooper pair breaks up in the left lead, with one electron tunneling
into the QD ($\propto\Gamma$) and one cotunneling into the quasiparticle
continuum above the gap in the right lead ($\propto\Gamma^{2}$).
Along the resonance $eV_{b}=-\frac{2}{3}E_{\downarrow}^{(0)}$ this
becomes energetically allowed when $eV_{b}>\frac{E_{g}}{3}$. For
voltages below this threshold the peak becomes very weak because although
the QD can be emptied by processes $\propto\Gamma^{3}$, higher order
processes are needed to fill it again. The peak increases further
in height for $eV_{b}>\frac{E_{g}}{2}$: here it becomes possible
to fill the QD by a sequential tunneling process where an electron
tunnels into the QD from the left lead ($\propto\Gamma$). An analogous
argument can be made to explain the lines with positive slope for
$E_{\downarrow}^{(0)}>0$, but with the roles of processes filling
and emptying the QD being reversed. An example of a higher-order process with $n=5$
is shown in Fig. \ref{MARMFQDMF}(g).
In addition to one Cooper pair cotunneling from the left to right lead through the QD (red and blue arrows, $\propto\Gamma^4$), one electron tunnels into the QD from the MBS (purple arrow, $\propto\Gamma$) which can only happen in the topological phase.

For $eV_{b}\to0$ and $E_{\downarrow}^{(0)}\to0$, the resonances
are seen to bend away from the linear voltage dependence described
by Eq.~(\ref{Mresonance2}). The reason is a level-repulsion effect
when the QD level comes close to the MBS.

In summary, the signatures of MBS are a series of unique straight
lines starting from $E_{\sigma}^{(0)}=0$ inside the gap according
to Eq.~(\ref{Mresonance2}).

Finally, we want to comment on the effect of Coulomb interactions between electrons 
on the QD (leading to Coulomb blockade),
which were neglected in this study. In the topological regime with MBS, the Zeeman
splitting is large and at low $V_b$ and close to $E_\downarrow^{(0)} = 0$, only
the spin-down state can be occupied even without Coulomb interactions. Therefore,
in this case we do not expect Coulomb interactions to drastically change the results
in the topological regime. In the opposite regime of small magnetic fields,
Coulomb interactions are expected to affect those resonances associated with double
occupation of the QD. Therefore, we would expect the signatures of Cooper pair tunneling
to change, whereas the MAR resonances described by Eq.~(\ref{Mresonance2})
[see Figs.~\ref{MARMFQDMF}(e)--(g)] should be less affected. We thus expect that
the qualitative difference of the stability diagram with and without MBS remains
also in the presence of strong Coulomb interactions. Nonetheless, including Coulomb
interactions is certainly an interesting problem for future studies.

\subsection{Transport in the topological phase with large tunnel coupling}

\begin{figure}
\centering\includegraphics[scale=0.53]{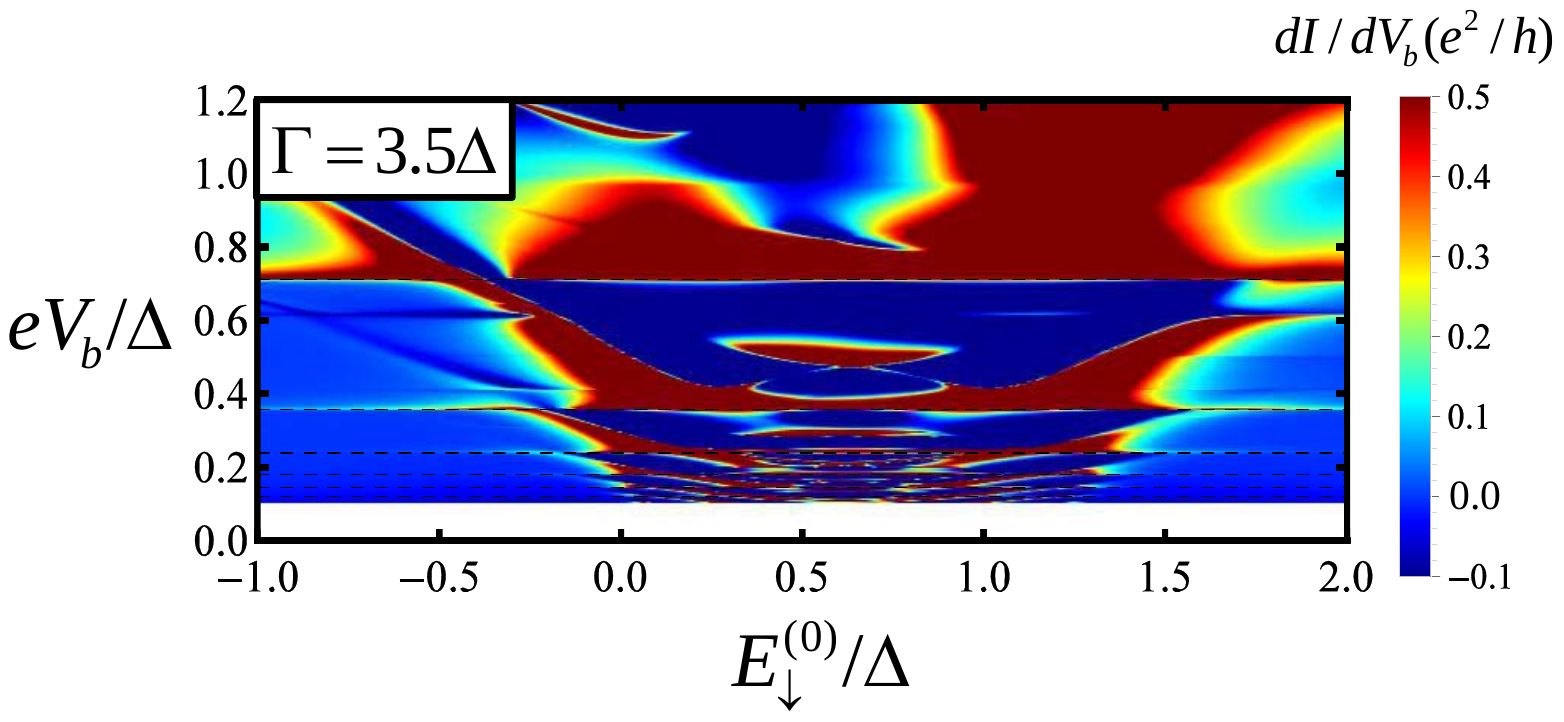}
\caption{\label{MARMFQDMFStrongCoupling} $dI/dV_{b}$ as a function of $eV_{b}$
and $E_{\downarrow}^{(0)}$ for the MBS-QD-MBS configuration with
large coupling $\Gamma=3.5\Delta$, where $V_{z}=2\Delta$ and $\mu=0$.
The horizontal dashed lines are guides to the eye indicating MAR peaks
at $eV_{b}=\frac{E_{g}}{m}$, where$E_{g}\approx0.71\Delta$, and
$m=1,2,\ldots,6$. The color scale is limited to the range $[-0.1,0.5]e^{2}/h$.
The other parameters are the same as in Fig. \ref{MARMFQDMF}.}
\end{figure}

As described above, when there is a small tunnel coupling between
the QD and leads, we see a series of peaks in $dI/dV_{b}$, the positions
of which depend linearly on the applied voltage bias and QD level
position. When the tunnel coupling becomes larger, higher order MAR
processes become increasingly important and the perturbatively oriented
picture we relied on earlier is no longer valid. The Green's function
method used for the actual calculations is, however, still accurate
and we show results with a large tunnel coupling in Fig.~\ref{MARMFQDMFStrongCoupling}.

In contrast to the case of weak tunnel coupling in Fig. \ref{MARMFQDMF}(a),
the main effect of changing the position of the QD level is to change
the strength of the peaks in $dI/dV_{b}$, which persist at $eV_{b}=\frac{E_{g}}{m}$,
$m=1,2,\ldots$ indicated by horizontal dashed lines in Fig.~\ref{MARMFQDMFStrongCoupling}.
This is the same as the position of MAR resonances for a MBS-weak
link-MBS structure.\cite{njp15.075019} The large coupling reduces
the role of the QD and the capability of tuning the transport with
a gate voltage is limited.

\section{Conclusions}

We have theoretically investigated tunnel spectroscopy of a S-QD-S
structure using the NEGF method. The peaks inside the superconducing
gap were analyzed in detail, both when the leads are in the trivial
and in the topological phase. In addition to Cooper pair tunneling, there are two classes
of electron tunneling processes, one relevant for the trivial phase
and one occuring only in the topological phase, giving rise to peaks
in the differential conductance which can be distinguished based on
their voltage dependence. In short, in the trivial phase the peaks
are related to the gap edge, giving the straight lines described by
Eq. (\ref{resonance}), while in topological phase, the MBS support
single electron tunneling in the middle of the gap, rendering peaks
along the lines described by Eq. (\ref{Mresonance2}). Based on our
findings, we suggest a S-QD-S junction with a gate-tunable QD level
as a promising platform for detection of MBS. In contrast to standard
tunnel spectroscopy, the presence of MBS qualitatively changes the
whole stability diagram, giving rise to peaks with a voltage dependence
which cannot be explained without zero-energy states at the edges
of the superconducting leads.

\section*{Acknowledgments}

G.-Y. Huang is grateful to Dr. Xizhou Qin for discussions about the
numerical calculations and to Dr. M. T. Deng for explaining the relevant
experiments. This work was supported by the Swedish Research Council
(VR), the Swedish Foundation for Strategic Research (SSF), the Chinese
Scholarship Council (CSC), the National Basic Research Program of China (Grants No. 2012CB932703 and No. 2012CB932700), 
the National Natural Science Foundation of China (Grants No. 91221202, No. 91431303, and No. KDB201400005),
the Danish National Research Foundation, and the Danish Council for Independent Research, Natural Sciences. 
\appendix
\section{Derivation of the expression for the stationary current}
The current flowing into the left contact is
\begin{eqnarray}
 I_{L}(\tau)&=&-e\langle\dot{N_{L}}\rangle\nonumber \\
 & = & -i\frac{e}{\hbar} \langle[H_{total},N_{L}]\rangle\nonumber \\
 & = & \frac{e}{\hbar}\mathrm{Re}\sum_{k}\mathrm{Tr}[G_{QD,Lk}^{<}(\tau,\tau)T_{L}(\tau)\tau_{z}],
\end{eqnarray}
where the electron number operator $N_{L}=\sum_{k\sigma}c_{Lk\sigma}^{\dagger}c_{Lk\sigma}$
and the total Hamiltonian is $H_{total}=\sum_{\alpha=L,R}H_{\alpha}+H_{c}+H_{QD}$,
where the mixed $4\times4$ Nambu Green's function $G_{QD,Lk}^{<}(\tau_{1},\tau_{2})$
is defined as
\begin{equation}
G_{QD,Lk}^{<}(\tau_{1},\tau_{2})=i\langle C_{Lk}^{\dagger}(\tau_{1})D(\tau_{2})\rangle.
\end{equation}
This Green's function can be found from the QD Green's function $G_{QD}^{r/<}(\tau_{1},\tau_{2})$
and lead Green's function $g_{Lk}^{a/<}(\tau_{1},\tau_{2})$ (the
derivation can be found in Ref. \cite{prb50.5528}, Appendix B)\widetext
\begin{equation}
G_{QD,Lk}^{<}(\tau,\tau)=\int_{-\infty}^{+\infty}d\tau'[G_{QD}^{r}(\tau,\tau')T_{L}^{\dagger}(\tau')g_{Lk}^{<}(\tau',\tau)+G_{QD}^{<}(\tau,\tau')T_{L}^{\dagger}(\tau')g_{Lk}^{a}(\tau',\tau)].
\end{equation}
Substituting this into the current gives
\begin{equation}
I_{L}(\tau)=\frac{e}{\hbar}\mathrm{Re}\int_{-\infty}^{+\infty}d\tau'\mathrm{Tr}[G_{QD}^{r}(\tau,\tau')\Sigma_{L}^{<}(\tau',\tau)+G_{QD}^{<}(\tau,\tau')\Sigma_{L}^{a}(\tau',\tau)]\tau_{z},
\end{equation}
\endwidetext\noindent where the lesser/advanced self energy on the left side
is
\begin{eqnarray}
\Sigma_{L}^{</a}(\tau',\tau) & = & \sum_{k}T_{L}^{\dagger}(\tau')g_{Lk}^{</a}(\tau',\tau)T_{L}(\tau) \\
 & = & T_{L}^{\dagger}\sum_{k}g_{Lk}^{</a}(\tau'-\tau)T_{L} \\
 & = & \Sigma_{L}^{</a}(\tau'-\tau) \\
 & = & \int\frac{d\omega}{2\pi}e^{-i\omega(\tau'-\tau)}\Sigma_{L}^{</a}(\omega),
\end{eqnarray}
which only depends on time difference because $T_{L}$ is time independent.
This is a consequence of applying the bias voltage only to the right
lead, and the lesser/advanced self energy on the right side is therefore
different
\begin{equation}
\Sigma_{R}^{</a}(\tau',\tau)=\sum_{k}T_{R}^{\dagger}(\tau')g_{Rk}^{</a}(\tau',\tau)T_{R}(\tau).
\end{equation}
The current $I_{L}(\tau)$ is periodic with period $T=\frac{2\pi}{\omega_{V}}$,
where $\omega_{V}=2eV_{b}$. This allows it to be expressed as the
Fourier expansion
\begin{equation}
I_{L}(\tau)=\sum_{n}e^{in\omega_{V}\tau}I_{n}.
\end{equation}
We will also need the Fourier expansion for the QD Green's function
and self energies
\begin{eqnarray}
G_{QD}(\tau_{1},\tau_{2}) & = & \sum_{n}e^{in\omega_{V}\tau_{1}}\int\frac{d\epsilon}{2\pi}e^{-i\epsilon(\tau_{1}-\tau_{2})}G_{QD,n}(\epsilon),\nonumber\\ \\
\Sigma_{L}(\tau_{1},\tau_{2}) & = & \int\frac{d\epsilon}{2\pi}e^{-i\epsilon(\tau_{1}-\tau_{2})}\Sigma_{L}(\epsilon),
\end{eqnarray}
where we neglected the superscripts because the relation holds for all self energies and QD Green's functions ($r,a,<,>$).
The Fourier component $I_{n}$ is\widetext
\begin{eqnarray}
I_{n} & = & \frac{1}{T}\int_{0}^{T}d\tau e^{-in\omega_{V}\tau}I_{L}(\tau) \\
 & = & \frac{1}{T}\int_{0}^{T}d\tau e^{-in\omega_{V}\tau}\frac{e}{\hbar}\mathrm{Re}\int_{-\infty}^{+\infty}d\tau'\mathrm{Tr}[G_{QD}^{r}(\tau,\tau')\Sigma_{L}^{<}(\tau',\tau)+G_{QD}^{<}(\tau,\tau')\Sigma_{L}^{a}(\tau',\tau)]\tau_{z} \\
 & = & \frac{e}{\hbar}\mathrm{Re}\mathrm{Tr}\frac{1}{T}\int_{0}^{T}d\tau e^{-in\omega_{V}\tau}\int_{-\infty}^{+\infty}d\tau'\sum_{m}e^{im\omega_{V}\tau}\int\frac{d\epsilon_{1}}{2\pi}e^{-i\epsilon_{1}(\tau-\tau')}\int\frac{d\epsilon_{2}}{2\pi}e^{-i\epsilon_{2}(\tau'-\tau)}G_{QD,m}(\epsilon_{1})\Sigma_{L}(\epsilon_{2})\tau_{z} \\
 & = & \frac{e}{\hbar}\mathrm{Re}\mathrm{Tr}\int\frac{d\epsilon_{1}}{2\pi}G_{QD,n}(\epsilon_{1})\Sigma_{L}(\epsilon_{1})\tau_{z},
\end{eqnarray}
\endwidetext\noindent where we used the shorthand notation $G_{QD}\Sigma_{L}=G_{QD}^{r}\Sigma_{L}^{<}+G_{QD}^{<}\Sigma_{L}^{a}$.
Following Ref.~\cite{prb78.024518} we rewrite this expression as
a sum of integrals over the fundamental domain $\omega\in F=[-\frac{\omega_{V}}{2},\frac{\omega_{V}}{2}]$,
\begin{eqnarray}
I_{n} & = & \frac{e}{\hbar}\mathrm{Re}\mathrm{Tr}\sum_{j=-\infty}^{\infty}\int_{j\omega_{V}-\frac{\omega_{V}}{2}}^{j\omega_{V}+\frac{\omega_{V}}{2}}\frac{d\epsilon}{2\pi}G_{QD,n}(\epsilon)\Sigma_{L}(\epsilon)\tau_{z} \nonumber \\ \\
 & = & \frac{e}{\hbar}\mathrm{Re}\mathrm{Tr}\sum_{j=-\infty}^{\infty}\int_{F}\frac{d\omega}{2\pi}G_{QD,n}(\omega_{j})\Sigma_{L}(\omega_{j})\tau_{z}\\
 & = & \frac{e}{\hbar}\mathrm{Re}\mathrm{Tr}\sum_{j=-\infty}^{\infty}\int_{F}\frac{d\omega}{2\pi}G_{QD,n+j,j}(\omega)\Sigma_{L,jj}(\omega)\tau_{z},\nonumber \\
\end{eqnarray}
where the shorthand notation $\omega_{j}=\omega+j\omega_{V}$ is used,
and $G_{QD,n+j,j}^{r/<}(\omega)\equiv G_{QD,n}^{r/<}(\omega_{j})$,
and the same for $\Sigma_{L,ij}^{a/<}(\omega)$. The time-averaged
current $I_{DC}$ is the zero harmonic
\begin{eqnarray}
 I_{DC} &=& I_{0} \\
 & = & \frac{e}{\hbar}\mathrm{Re}\mathrm{Tr}\sum_{j=-\infty}^{\infty}\int_{F}\frac{d\omega}{2\pi}G_{QD,jj}(\omega)\Sigma_{L,jj}(\omega)\tau_{z} \nonumber \\ \\
 & = & \frac{e}{\hbar}\mathrm{Re}\mathrm{Tr}\int_{F}\frac{d\omega}{2\pi}[\boldsymbol{G}_{QD}(\omega)\boldsymbol{\Sigma}_{L}(\omega)\tau_{z}] \\
 & = & \frac{e}{h}\mathrm{Re}\mathrm{Tr}\int_{F}d\omega \nonumber \\
 &\times& [\boldsymbol{G}_{QD}^{r}(\omega)\boldsymbol{\Sigma}_{L}^{<}(\omega)+\boldsymbol{G}_{QD}^{<}(\omega)\boldsymbol{\Sigma}_{L}^{a}(\omega)]\tau_{z},
\end{eqnarray}
where $\boldsymbol{G}$ and $\boldsymbol{\Sigma}$ are matrices
in terms of the Fourier components and the third equality follows from the fact that
the trace also operates on Fourier space and $\boldsymbol{\Sigma}$
is diagonal [Eq.~(\ref{eqSelfEnergyL})]. The last equality expands
the shorthand notation. This is the expression used in the main text.

The current above is straightforwardly constructed starting from the numerically calculated Green's functions (see the equations in Sec. II). A check of the convergence of the current will help us to confirm the reliability of our results. The current including $K$ Fourier components is
\begin{eqnarray}
 I_{DC,K} &=& \frac{e}{\hbar}\mathrm{Re}\mathrm{Tr}\sum_{j=-K}^{K}\int_{F}\frac{d\omega}{2\pi}G_{QD,jj}(\omega)\Sigma_{L,jj}(\omega)\tau_{z} \nonumber \\ \\
  & = & \frac{e}{h}\mathrm{Re}\mathrm{Tr}\int_{F}d\omega \nonumber \\
 &\times& [\boldsymbol{G}_{QD}^{r}(\omega)\boldsymbol{\Sigma}_{L}^{<}(\omega)+\boldsymbol{G}_{QD}^{<}(\omega)\boldsymbol{\Sigma}_{L}^{a}(\omega)]\tau_{z},
\end{eqnarray}
where the size of the matrices $\boldsymbol{G}_{QD}^{r,a,<}(\omega),\boldsymbol{\Sigma}_{L,R}^{r,a,<}(\omega)$ is $4(2K+1)\times4(2K+1)$.
When the number $K$ of Fourier components increases, the current will converge, $I_{DC,converged}=\lim_{K\rightarrow \infty}I_{DC,K}$.

\begin{figure}[htb!]
\centering\includegraphics[scale=0.5]{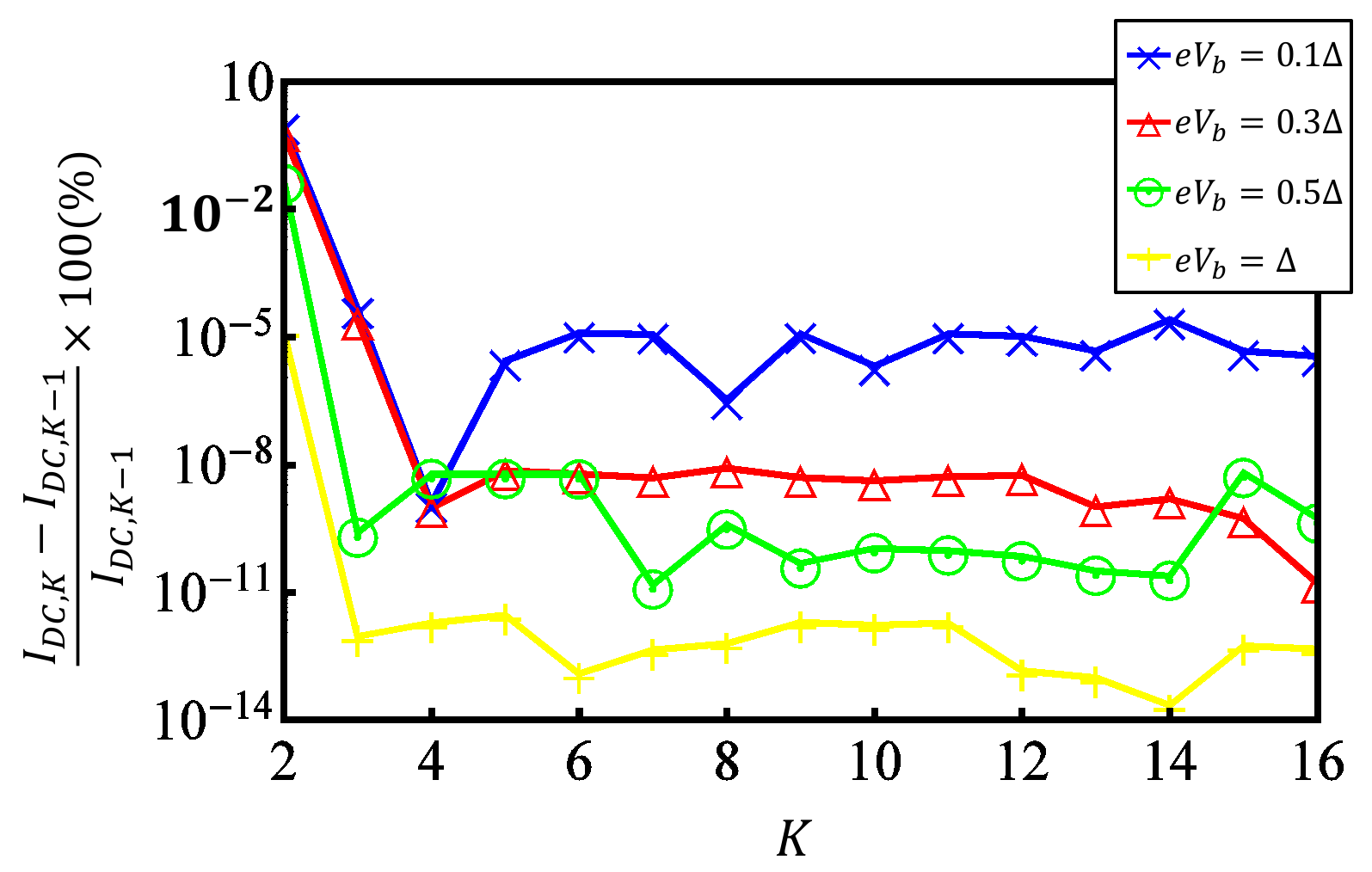}
\caption{\label{convergenceCheck} A convergence check of the current for some example points along the line $E_\downarrow^{(0)}=0.5\Delta$ in Fig.~\ref{MARTrivial}(a) with an increasing number $K$ of Fourier components included. The relative error function is $(I_{DC,K}-I_{DC,K-1})/I_{DC,K-1}\times 100(\%)$.}
\end{figure}

In practical calculation, an adaptive algorithm is used, where $K$ is increased until a certain error condition is met. In Fig.~\ref{convergenceCheck} we plot the relative error $(I_{DC,K}-I_{DC,K-1})/I_{DC,K-1}\times 100(\%)$. For large bias (roughly $eV_b\geq 0.1\Delta$ in our case), the current will be converged after only a few steps.


\begin{thebibliography}{10}
\bibitem{np5.614}F. Wilczek, Nature Phys. {\bf 5}, 614 (2009).

\bibitem{rmp80.1083}C. Nayak, S. H. Simon, A. Stern, M. Freedman, and S. Das Sarma, Rev. Mod. Phys. {\bf 80}, 1083 (2008).

\bibitem{prb61.10267}N. Read and D. Green, Phys. Rev. B {\bf 61}, 10267 (2000).

\bibitem{prl86.268}D. A. Ivanov, Phys. Rev. Lett. {\bf 86}, 268 (2001).

\bibitem{prl100.096407}L. Fu and C. L. Kane, Phys. Rev. Lett. {\bf 100}, 096407 (2008).

\bibitem{prl104.040502}J. D. Sau, R. M. Lutchyn, S. Tewari, and S. Das Sarma, Phys. Rev. Lett. {\bf 104}, 040502 (2010).

\bibitem{prb82.214509}J. D. Sau, S. Tewari, R. Lutchyn, T. Stanescu, and S. Das Sarma, Phys. Rev. B {\bf 82}, 214509 (2010).

\bibitem{prl105.077001}R. M. Lutchyn, J. D. Sau, and S. Das Sarma, Phys. Rev. Lett. {\bf 105}, 077001 (2010).

\bibitem{prl105.177002}Y. Oreg, G. Refael, and F. von Oppen, Phys. Rev. Lett. {\bf 105}, 177002 (2010).

\bibitem{rpp75.076501}J. Alicea, Rep. Prog. Phys. {\bf 75}, 076501 (2012).

\bibitem{sst27.124003}Martin Leijnse and Karsten Flensberg, Semicond. Sci. Technol. {\bf 27}, 124003 (2012).

\bibitem{arcmp4.113}C. W. J. Beenakker, Annu. Rev. Condens. Matter Phys. {\bf 4}, 113 (2013).

\bibitem{science336.1003}V. Mourik, K. Zuo, S. M. Frolov, S. R. Plissard, E. P. A. M. Bakkers, and L. P. Kouwenhoven, Science {\bf 336}, 1003 (2012).

\bibitem{np8.887}Anindya Das, Yuval Ronen, Yonatan Most, Yuval Oreg, Moty Heiblum, and Hadas Shtrikman, Nature Phys. {\bf 8}, 887 (2012).

\bibitem{nl12.6414}M. T. Deng, C. L. Yu, G. Y. Huang, M. Larsson, P. Caroff, and H. Q. Xu, Nano Lett. {\bf 12}, 6414 (2012).

\bibitem{arXiv1406.4435}M. T. Deng, C. L. Yu, G. Y. Huang, M. Larsson, P. Caroff, and H. Q. Xu, Sci. Rep. {\bf 4}, 7261 (2014).

\bibitem{prb87.241401}H. O. H. Churchill, V. Fatemi, K. Grove-Rasmussen, M. T. Deng, P. Caroff, H. Q. Xu, and C. M. Marcus, Phys. Rev. B {\bf 87}, 241401 (2013).

\bibitem{prl110.126406}A. D. K. Finck, D. J. Van Harlingen, P. K. Mohseni, K. Jung, and X. Li, Phys. Rev. Lett. {\bf 110}, 126406 (2013).

\bibitem{prl109.186802}Eduardo J. H. Lee, Xiaocheng Jiang, Ram\'{o}n Aguado, Georgios Katsaros, Charles M. Lieber, and Silvano De Franceschi, Phys. Rev. Lett. {\bf 109}, 186802 (2012).

\bibitem{nnano9.79}Eduardo J. H. Lee, Xiaocheng Jiang, Manuel Houzet, Ram\'{o}n Aguado, Charles M. Lieber, and Silvano De Franceschi, Nature Nanotech. {\bf 9}, 79 (2014).

\bibitem{njp14.125011}D. I. Pikulin, J. P. Dahlhaus, M. Wimmer, H. Schomerus, C. W. J. Beenakker, New J. Phys. {\bf 14}, 125011 (2012).

\bibitem{prl110.217005}W. Chang, V. E. Manucharyan, T. S. Jespersen, J. Nyg\aa{}rd, and C. M. Marcus, Phys. Rev. Lett. {\bf 110}, 217005 (2013).

\bibitem{prb85.060507} G. Kells, D. Meidan, and P. W. Brouwer, Phys. Rev. B {\bf 85}, 060507(R) (2012).

\bibitem{prl109.267002}Jie Liu, Andrew C. Potter, K.T. Law, and Patrick A. Lee, Phys. Rev. Lett. {\bf 109}, 267002 (2012).

\bibitem{pu44.131}A. Y. Kitaev, Phys. Usp. {\bf 44}, 131 (2001).

\bibitem{prl109.056803}J. R. Williams, A. J. Bestwick, P. Gallagher, Seung Sae Hong, Y. Cui, Andrew S. Bleich, J. G. Analytis, I. R. Fisher, and D. Goldhaber-Gordon, Phys. Rev. Lett. {\bf 109}, 056803 (2012).

\bibitem{np8.795}Leonid P. Rokhinson, Xinyu Liu, and Jacek K. Furdyna, Nature Phys. {\bf 8}, 795 (2012).

\bibitem{Badiane11} D. M. Badiane, M. Houzet, and J. S. Meyer, Phys. Rev. Lett. {\bf 107}, 177002 (2011).

\bibitem{njp15.075019}Pablo San-Jose, Jorge Cayao, Elsa Prada and Ram\'{o}n Aguado, New J. Phys. {\bf 15}, 075019 (2013).

\bibitem{science309.272} Y.-J. Doh, J. A. {van Dam}, A. L. Roest, E. P. A. M. Bakkers, L. P. Kouwenhoven and S. {De Franceschi}, Science {\bf 309}, 272 (2005).

\bibitem{nature442.667} J. A. {van Dam}, Y. V. Nazarov, E. P. A. M. Bakkers, S. {De Franceschi} and L. P. Kouwenhoven, Nature {\bf 442}, 667 (2006).

\bibitem{nl12.228}H. A. Nilsson, P. Samuelsson, P. Caroff, and H. Q. Xu, Nano Lett. {\bf 12}, 228 (2012).

\bibitem{nl12.5622} S. Abay, H. Nilsson, F. Wu, H. Q. Xu, C. M. Wilson, and P. Delsing, Nano Lett. {\bf 12}, 5622 (2012).

\bibitem{nl13.3614}S. Abay, D. Persson, H. Nilsson, H. Q. Xu, M. Fogelstr\"{o}m, V. Shumeiko, and P. Delsing, Nano Lett. {\bf 13}, 3614 (2013).

\bibitem{prb89.214508}S. Abay, D. Persson, H. Nilsson, F. Wu, H. Q. Xu, M. Fogelstr\"{o}m, V. Shumeiko, and P. Delsing, Phys. Rev. B {\bf 89}, 214508 (2014).

\bibitem{nl9.3151}H. A. Nilsson, P. Caroff, C. Thelander, M. Larsson, J. B. Wagner, L.-E. Wernersson, L. Samuelson, and H. Q. Xu, Nano Lett. {\bf 9}, 3151 (2009).

\bibitem{prb61.4754}Qing-feng Sun, Bai-geng Wang, Jian Wang, and Tsung-han Lin, Phys. Rev. B {\bf 61}, 4754 (2000).

\bibitem{prl107.256802}B. M. Andersen, K. Flensberg, V. Koerting, and J. Paaske, Phys. Rev. Lett. {\bf 107}, 256802 (2011).

\bibitem{prb66.165305}H. Q. Xu, Phys. Rev. B {\bf 66}, 165305 (2002).

\bibitem{prb72.195346}F. Zhai and H. Q. Xu, Phys. Rev. B {\bf 72}, 195346 (2005).

\bibitem{Wimmer}M. Wimmer, PhD Thesis (Universit\"at Regensburg, 2008), Chapter 3 and Appendix D.

\bibitem{prb78.024518}Fabrizio Dolcini, and Luca Dell'Anna, Phys. Rev. B {\bf 78}, 024518 (2008).

\bibitem{prb27.6739}M. Octavio, M. Tinkham, G. E. Blonder, and T. M. Klapwijk, Phys. Rev. B {\bf 27}, 6739 (1983).

\bibitem{prb38.8707}K. Flensberg, J.~B. Hansen, and M. Octavio, Phys. Rev. B {\bf 38}, 8707 (1988).

\bibitem{prl10.17}J.~R. Schrieffer J.~W. Wilkins Phys. Rev. Lett. {\bf 10}, 17 (1963).

\bibitem{prb9.3757}L.~E. Hasselberg, M.~T. Levinsen, and M.~R. Samuelsen, Phys. Rev. B {\bf 9}, 3757 (1974).

\bibitem{jltp68.1}G.~B. Arnold, J. Low Temp. Phys. {\bf 68}, 1 (1987).

\bibitem{prl74.2110}E. N. Bratus, V. S. Shumeiko, and G. Wendin, Phys. Rev. Lett. {\bf 74}, 2110 (1995).

\bibitem{prb54.7366}J. C. Cuevas, A. Mart\'{i}n-Rodero, and A. Levy Yeyati, Phys. Rev. B {\bf 54}, 7366 (1996).

\bibitem{prl75.1831}D. Averin and A. Bardas, Phys. Rev. Lett. {\bf 75}, 1831 (1995).

\bibitem{prb64.144504}\AA . Ingerman, G. Johansson, V. S. Shumeiko, and G. Wendin, Phys. Rev. B {\bf 64}, 144504 (2001).

\bibitem{prb60.1382}G. Johansson, E. N. Bratus, V. S. Shumeiko, and G. Wendin, Phys. Rev. B {\bf 60}, 1382 (1999).

\bibitem{prb73.214501}A. Zazunov, R. Egger, C. Mora, and T. Martin, Phys. Rev. B {\bf 73}, 214501 (2006).

\bibitem{prb55.6137}A. Levy Yeyati, J. C. Cuevas, A. L\'{o}pez-D\'{a}valos, and A. Mart\'{i}n-Rodero, Phys. Rev. B {\bf 55}, R6137 (1997).

\bibitem{prb65.075315}Qing-feng Sun, Hong Guo, and Jian Wang, Phys. Rev. B {\bf 65}, 075315 (2002).

\bibitem{prb80.184510}T. Jonckheere, A. Zazunov, K. V. Bayandin, V. Shumeiko, and T. Martin, Phys. Rev. B {\bf 80}, 184510 (2009).

\bibitem{prl91.057005}M. R. Buitelaar, W. Belzig, T. Nussbaumer, B. Babi\'{c}, C. Bruder, and C. Sch\"{o}nenberger, Phys. Rev. Lett. {\bf 91}, 057005 (2003).

\bibitem{prb65.180514}P. Samuelsson, G. Johansson, \AA . Ingerman, V. S. Shumeiko, and G. Wendin, Phys. Rev. B {\bf 65}, 180514(R) (2002).

\bibitem{prb86.235427} B. Hiltscher, M. Governale, and J. K\"onig, Phys. Rev. B {\bf 86}, 235427 (2012).

\bibitem{prb77.161406}J. V. Holm, H. I. J\o{}rgensen, K. Grove-Rasmussen, J. Paaske, K. Flensberg, and P. E. Lindelof, Phys. Rev. B {\bf 77}, 161406(R) (2008).

\bibitem{prb50.5528}Antti-Pekka Jauho, Ned S. Wingreen, and Yigal Meir, Phys. Rev. B {\bf 50}, 5528 (1994).\end{thebibliography}
\end{document}